\documentclass[]{aastex701}

\newcommand{\hi}{H\,\textsc{i}}
\newcommand{\kms}{km\,s$^{-1}$}
\newcommand{\msun}{M_\odot}
\newcommand{\ha}{H$\alpha$}

\newcommand{\nii}{[N\,\textsc{ii}]}
\newcommand{\sii}{[S\,\textsc{ii}]}
\newcommand{\oi}{[O\,\textsc{i}]}
\newcommand{\oiii}{[O\,\textsc{iii}]}

\shorttitle{Counter-rotating gas in a star-forming elliptical}
\shortauthors{Biswas et al.}

\begin{document}

\title{Counter-rotating Ionized and Neutral Gas in a Star-forming Elliptical Galaxy}

\author[0009-0002-2173-0953]{Pralay Biswas}
\affiliation{National Centre for Radio Astrophysics, Tata Institute of Fundamental Research, Post Bag 3, Ganeshkhind, Pune 411007, India}
\email[show]{pbiswas@ncra.tifr.res.in}
\correspondingauthor{Pralay Biswas}

\author[0000-0002-1345-7371]{Yogesh Wadadekar}
\affiliation{National Centre for Radio Astrophysics, Tata Institute of Fundamental Research, Post Bag 3, Ganeshkhind, Pune 411007, India}
\email{yogesh@ncra.tifr.res.in}

\author[0000-0003-2722-8841]{Omkar Bait}
\affiliation{National Radio Astronomy Observatory, 520 Edgemont Road, Charlottesville, VA 22903, USA}
\affiliation{The NSF-Simons AI Institute for Cosmic Origins, USA, 201 E. 24th Street, POB 4.102, Austin, Texas 78712-1229}
\email{obait@nrao.edu}

\author[0000-0001-6615-5492]{Sushma Kurapati}
\affiliation{National Centre for Radio Astrophysics, Tata Institute of Fundamental Research, Post Bag 3, Ganeshkhind, Pune 411007, India}
\email{sushma@ncra.tifr.res.in}

\begin{abstract}
In early-type galaxies, gas that is kinematically misaligned with the stars generally rotates together across all gas phases, implying a single external origin. Using upgraded Giant Metrewave Radio Telescope (uGMRT) 21 cm imaging and SDSS-MaNGA spatially resolved spectroscopy, we studied MaNGA~10214-1902, a compact ($R_e = 0.78$ kpc), star-forming elliptical galaxy with stellar mass $\log M_*/\msun = 9.90$, in which the ionized gas counter-rotates with respect to the stars. It hosts $\log(M_{\rm HI}/\msun) = 9.65$ in a 45 kpc \hi\ disk that co-rotates with the stars and has a central column-density depression. Inside this depression, the ionized gas ($r \lesssim 2$ kpc) rotates in the opposite sense to both the stars ($\Delta{\rm PA} = 152^\circ \pm 20^\circ$) and the \hi\ that surrounds it ($\Delta{\rm PA} = 179^\circ \pm 10^\circ$). \hi\ co-spatial with the ionized gas is not detected, so a fainter inner \hi\ component with a different orientation cannot be ruled out. This central gas is star-forming in the \sii\ and \oi\ diagnostic diagrams and has EW(\ha)\,$>6$\,\AA\ at the center. The 1.4 GHz radio-continuum star formation rate agrees with the spectral energy distribution based value, showing no evidence of a dominant active galactic nucleus. The stellar population at the center is younger than the outskirts and has the lowest D$_{\rm n}$4000. These results are consistent with the acquisition of gas with opposite angular momentum into the center of a pre-existing \hi\ disk that co-rotates with the stars, which can fuel the most recent central star formation.
\end{abstract}

\keywords{\uat{Elliptical galaxies}{456} --- \uat{Early-type galaxies}{429} --- \uat{Galaxy kinematics}{602} --- \uat{Interstellar atomic gas}{833} --- \uat{HI line emission}{690} --- \uat{Star formation}{1569} --- \uat{Galaxy evolution}{594}}

\section{Introduction} \label{sec:intro}

Galaxies in the local universe show a bimodal distribution of actively star-forming (blue cloud) galaxies that follow the star-forming main sequence (SFMS) and passively evolving (red sequence) galaxies \citep{Strateva2001, Baldry2004, Brinchmann2004, Salim2007}. Most early-type galaxies (ETGs; ellipticals and lenticulars) are red, gas-poor, and have little to no star formation \citep{Kauffmann2003may, Bell2004}. However, a small number of ETGs are actively forming stars and lie on the SFMS \citep{Schawinski2009, George2015, George2017, Rathore2022}. In a sample of 59 such star-forming ellipticals (SF-Es) drawn from SDSS-MaNGA, \citet{Biswas2024} found that most of them have kinematically disturbed stellar and ionized gas velocity maps and a mixture of young and old stellar populations, which indicates that they have acquired fresh gas through recent minor mergers or gas accretion. So, to directly identify the acquired gas, resolved observations of the cold gas are necessary.

Neutral atomic hydrogen (\hi) is the dominant reservoir of cold gas in galaxies and ultimately fuels star formation \citep{Elmegreen1993}. Although ETGs are traditionally viewed as \hi-poor, a substantial fraction of them host \hi, often in large, regularly rotating disks or rings \citep{Morganti2006, Oosterloo2010, Serra2012, Bait2020}. Kinematic misalignment between the gas and the stars is one of the cleanest signatures of such externally acquired gas, because the gas recycled internally from evolved stars inherits the angular momentum of the stars \citep{Bertola1992}. In the ATLAS$^{3D}$ survey, $36\pm5\%$ of fast-rotating ETGs have ionized gas misaligned with the stars \citep{Davis2011}. Studies using SDSS-MaNGA have found misaligned and counter-rotating gas in quiescent, green valley, and also in a small fraction of blue star-forming galaxies \citep{Jin2016, Chen2016, Zhou2022, Rathore2022}, and in the SAMI survey, $\approx11\%$--12\% of all galaxies and $\approx45\%$ of ETGs have gas misaligned with the stars \citep{Bryant2019, Ristea2022}. Misaligned gas has also been linked to enhanced black hole activity \citep{Raimundo2023}. In the star-forming counter-rotators, the collisions between the accreted and the pre-existing gas remove angular momentum, funnel the gas toward the center, and grow the central region \citep{Chen2016}, and this channel can sustain main-sequence star formation below $M_* \sim 10^{10}\,\msun$ \citep{Tous2026}.

Notably, when the gas in an ETG is misaligned with the stars, its atomic, molecular, and ionized phases are found to be kinematically aligned with each other where they overlap, which implies that all three phases share a common origin \citep{Davis2011}. Galaxies with two misaligned gas components are rare; fewer than ten are known, mostly from long-slit spectroscopy \citep{Rubin1994, Fisher1994, Plana1996, Hunter1999, Corsini2002}, including the spiral galaxy NGC~4826 with two nested, counter-rotating gaseous disks \citep{Braun1992, Braun1994} and the dwarf irregular galaxy NGC~4449 with counter-rotating inner and outer \hi\ \citep{Hunter1999}. Recently, \citet{cao2022Nat} found two MaNGA galaxies with dual misaligned ionized gas disks, implying multiple gas acquisition events. However, an ETG with counter-rotating ionized gas at the center of a \hi\ disk co-rotating with the stars has not been reported.

In this Letter, we present uGMRT \hi\ and SDSS-MaNGA observations of MaNGA~10214-1902, an SF-E galaxy that shows this configuration. Throughout, we use the $\Lambda$CDM cosmology with $H_0 = 67.7$\,\kms\,Mpc$^{-1}$, $\Omega_m = 0.31$, and $\Omega_\Lambda = 0.69$ \citep[Planck18;][]{Planck2020}, for which the luminosity distance of the target ($z = 0.02178$) is 98.1\,Mpc and $1''$ corresponds to 0.456\,kpc.

\section{Target, Observations, and Data} \label{sec:data}

\subsection{Target Selection} \label{sec:target}

MaNGA~10214-1902 is one of the 59 SF-Es of \citet{Biswas2024}. It was selected for \hi\ follow-up because its MaNGA velocity maps already showed that the ionized gas counter-rotates with respect to the stars, and its single-dish \hi\ spectrum from the \hi-MaNGA survey with the Green Bank Telescope \citep[GBT;][]{Masters2019, Stark2021} showed that it is \hi-rich. The deep-learning morphology of \citet{Dominguez2022} classifies this galaxy as an elliptical (T-Type $=-3.1$), whereas the visual classification of \citet{VazquezMata2022} assigns an S0 type and flags tidal features. All the global properties of the galaxy are listed in Table~\ref{tab:props}.

\subsection{Observation and Ancillary Data} \label{sec:obs}

MaNGA~10214-1902 was observed with the uGMRT \citep{Gupta2017} Band~5 receiver and the GMRT Wideband Backend \citep{Reddy2017} on 2025 January 5 for $\approx9$ hr of on-source time. The data were calibrated, self-calibrated, and continuum-subtracted in \texttt{CASA} \citep{casa}. The \hi\ line emission in individual channels and the corresponding moment maps, and the integrated \hi\ spectrum, were derived with SoFiA-2 \citep{Serra2015, Westmeier2021} as described in Appendix~\ref{app:gmrt}. The GMRT has both a compact central array and long baselines, which allowed us to produce six spectral cubes with 10\,\kms\ velocity resolution and angular resolutions ranging from $3.3''\times2.8''$ to $43''\times36''$ (Table~\ref{tab:cubes}). Hereafter, we refer to each cube by its angular resolution. We use the $28.9''\times25.3''$ cube ($0.5$--$10\,{\rm k}\lambda$) for the \hi\ morphology, mass, and size, and the $11.3''\times9.7''$, $5.9''\times5.2''$, and $3.3''\times2.8''$ cubes ($0.5$--$20$, $0.5$--$40$, and $0.5$--$80\,{\rm k}\lambda$) only to get high column density \hi\ distribution and kinematics in higher spatial resolution (Appendix~\ref{app:gmrt}).

Optical integral field spectroscopy of the target is available from the SDSS-IV MaNGA survey \citep{Bundy2015, Drory2015, LawMaNGAdrp2016, Wake2017, Abdurrouf2022}, observed with a 19-fiber integral field unit (IFU) of $12.5''$ diameter and an $r$-band point-spread function (PSF) of $2.7''$ full width at half maximum (FWHM). All the spatially resolved quantities, the stellar and ionized gas kinematics, the emission line fluxes and equivalent widths, and the D$_{\rm n}$4000 index \citep{Balogh1999} are taken from the DR17 MaNGA Data Analysis Pipeline \citep[DAP;][]{WestfallMaNGAdap2019, BelfioreMaNGAdap2019} \texttt{VOR10-MILESHC-MASTARSSP} maps. We used S/N\,$>5$ for the kinematics and S/N\,$>3$ in all the lines used in the diagnostic diagrams. As the DAP does not provide the stellar population ages, only the light- and mass-weighted ages and the light fractions in different age bins are taken from Pipe3D \citep{SanchezI2016, SanchezII2016, SanchezPipe3Ddr2022}. Here, we note that the effective radius of the target ($R_e = 1.71''$, the NSA\footnote{NASA-Sloan Atlas  (\url{https://nsatlas.org/})} elliptical Petrosian half-light radius) is smaller than the PSF, and therefore the kinematics within $R_e$ are not resolved.

We use the GALEX-SDSS-WISE Legacy Catalog \citep[GSWLC-A2;][]{Salim2016, Salim2018}, hereafter the Salim catalog, for obtaining stellar mass, star formation rate (SFR), and dust attenuation. The Salim catalog derives these quantities by modeling the ultraviolet-to-infrared spectral energy distribution (SED) with the CIGALE code \citep{Noll2009, Boquien2019}, using photometry from GALEX, SDSS, and WISE. For the $r$-band S\'ersic index, we use the MaNGA PyMorph DR17 catalog \citep{Vikram2010, Fischer2019, Dominguez2022}. In the second data release of \hi-MaNGA, all the fluxes obtained with GBT were multiplied by 1.2 to correct the flux calibration scale of the reduction software GBTIDL\footnote{\url{https://gbtidl.nrao.edu/}} \citep{Stark2021}. We have applied this factor to the archived GBT spectrum. No source confusion is flagged within the $\sim 9'$ GBT beam (the blueberry galaxy J0807+3414 at a similar redshift is not detected in \hi; Appendix~\ref{app:j0807}), and the bright galaxy $\sim1'$ west of the target is a background galaxy at $z = 0.096$. The optical images are obtained from the DESI Legacy Imaging Surveys \citep{Dey2019}.

\begin{deluxetable*}{lcl}
\tablecaption{Properties of MaNGA 10214-1902 \label{tab:props}}
\tablehead{\colhead{Quantity} & \colhead{Value} & \colhead{Source}}
\startdata
\multicolumn{3}{c}{\textit{Optical}} \\
R.A., Decl. (J2000)                          & 08:07:57.19, +34:13:38.4 & NSA \\
Redshift, $z$                                & 0.02178                  & NSA \\
$\log(M_*/\msun)$                            & $9.90\pm0.02$            & Salim catalog \\
$\log({\rm SFR}/\msun\,{\rm yr^{-1}})$       & $-0.65\pm0.14$           & Salim catalog \\
$\log({\rm sSFR/yr^{-1}})$                   & $-10.55$                 & Salim catalog \\
$A_V$ (mag)                                  & $0.33\pm0.14$            & Salim catalog \\
$R_e$                                        & $1.71''$ (0.78\,kpc)     & NSA \\
S\'ersic index, $n_r$                        & $3.6$                    & PyMorph\tablenotemark{a} \\
D$_{\rm n}$4000, center / $3''$--$6''$       & 1.60 / 1.67              & DAP \\
EW(\ha), center (\AA)                        & 8.1                      & DAP \\
$\sigma_*$ within $R_e$ (\kms)               & 77                       & DAP \\
Light-/mass-weighted age within $R_e$ (Gyr)  & 1.9 / 3.5                & Pipe3D \\
Light-/mass-weighted age at $3''$--$6''$ (Gyr)  & 3.2 / 5.2              & Pipe3D \\
SFR$_{{\rm H}\alpha}$ ($\msun$\,yr$^{-1}$)  & 0.10     & This work \\
\hline
\multicolumn{3}{c}{\hi\ \textit{and radio continuum}} \\
$S_{\rm HI}$, uGMRT (Jy\,\kms)               & 1.64                     & This work \\
$F_{\rm HI}$, GBT (Jy\,\kms)                 & $2.06\pm0.09$            & \hi-MaNGA\tablenotemark{b} \\
$\log(M_{\rm HI}/\msun)$, uGMRT / GBT        & 9.55 / 9.65              & This work\tablenotemark{c} \\
$W_{50}$ (\kms)                              & 141                      & \hi-MaNGA \\
$D_{\rm HI}$ (kpc)                           & 45                       & This work\tablenotemark{d} \\
$D_{\rm HI}$ predicted (kpc)                 & 35--39                   & This work\tablenotemark{e} \\
\hi\ inclination (deg)                       & 43                       & This work \\
$S_{1.4\,{\rm GHz}}$ ($\mu$Jy)               & $284\pm38$               & This work \\
SFR$_{1.4\,{\rm GHz}}$ ($\msun$\,yr$^{-1}$)  & $0.21\pm0.03$            & This work \\
\hline
\multicolumn{3}{c}{\textit{Kinematic position angles (receding side, east of north)\tablenotemark{f}}} \\
Stars (DAP)                                  & $195^\circ\pm19^\circ$   & This work \\
\ha\ (DAP)                                   & $42^\circ\pm7^\circ$     & This work \\
\hi, $28.9''\times25.3''$              & $219^\circ\pm7^\circ$    & This work \\
\hi, $5.9''\times5.2''$, $r<12''$   & $224^\circ\pm7^\circ$    & This work \\
\enddata
\tablenotetext{a}{Single-S\'ersic fit, preferred by PyMorph (\texttt{FLAG\_FIT}\,$=1$); the S\'ersic+exponential fit gives $B/T = 0.72$.}
\tablenotetext{b}{Including the factor of 1.2 flux-scale correction of \hi-MaNGA DR2 \citep{Stark2021}.}
\tablenotetext{c}{Both at the Planck18 luminosity distance of 98.1\,Mpc. \hi-MaNGA lists $\log M_{\rm HI} = 9.62$ for $H_0 = 70$\,\kms\,Mpc$^{-1}$.}
\tablenotetext{d}{Major axis at $\Sigma_{\rm HI} = 1\,\msun$\,pc$^{-2}$, beam-corrected, measured following \citet{WangJ2016}.}
\tablenotetext{e}{Predicted from the \hi\ size--mass relation of \citet{WangJ2016} for the uGMRT and GBT \hi\ masses.}
\tablenotetext{f}{From \texttt{fit\_kinematic\_pa} \citep{Krajnovic2006}; the uncertainties are $3\sigma$.}
\end{deluxetable*}

\section{Results} \label{sec:results}

\subsection{\texorpdfstring{\hi\ Content and Distribution}{HI Content and Distribution}} \label{sec:hi}

\begin{figure*}[t!]
\centering
\includegraphics[width=\textwidth]{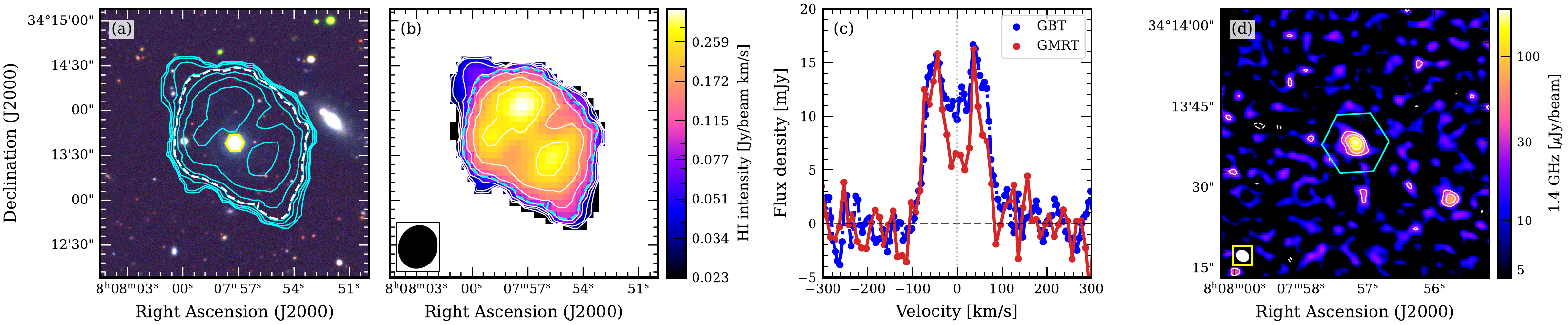}
\caption{(a) DESI Legacy Surveys $grz$ image with the uGMRT \hi\ column-density contours (cyan) from the $28.9''\times25.3''$ cube and the MaNGA IFU footprint (yellow hexagon). The contours start at $3\sigma$ ($N_{\rm HI} = 3.4\times10^{19}$\,cm$^{-2}$) and increase by factors of 1.5, and the dashed white contour marks $\Sigma_{\rm HI} = 1\,\msun$\,pc$^{-2}$. The bright galaxy to the west is a background galaxy at $z = 0.096$. (b) \hi\ moment-0 map with the same contours and spatial resolution. (c) \hi\ spectra from the GBT (HI-MaNGA, blue, on the DR2 flux scale) and the uGMRT (red), relative to $cz = 6544.5$\,\kms. (d) uGMRT 1.4 GHz continuum image with contours at $(\pm 1, 2, 4, \ldots)\times3\sigma$ ($\sigma = 9.0\,\mu$Jy\,beam$^{-1}$), the MaNGA hexagonal footprint (cyan), and the beam ($2.3''\times1.9''$; white ellipse in the yellow box).}
\label{fig:hi}
\end{figure*}

Figure~\ref{fig:hi} shows the \hi\ distribution and spectra of the target. The uGMRT imaging gives $S_{\rm HI} = 1.64$ Jy \kms, or $\log(M_{\rm HI}/\msun) = 9.55$, and the GBT spectrum gives $\log(M_{\rm HI}/\msun) = 9.65$. The \hi-to-stellar mass ratio is therefore $M_{\rm HI}/M_* \approx 0.5$, which is $\approx0.5$ dex above the average \hi\ fraction of galaxies with the same stellar mass in xGASS \citep{Catinella2018}. Both spectra are double-horned with $W_{50} = 141$\,\kms. The uGMRT recovers 73\%--79\% of the GBT flux; the missing flux is concentrated near the systemic velocity and in the high-velocity horn. This indicates that the missing flux may lie in extended, low-surface-brightness emission that the interferometer does not recover. This does not affect the disk size measurements below, which are obtained at a higher column density level.

Following \citet{WangJ2016}, we measured the \hi\ diameter, $D_{\rm HI}$, as the major axis of the ellipse where the azimuthally averaged surface density reaches $1\,\msun$\,pc$^{-2}$, corrected for the beam. We find $D_{\rm HI} = 45$ kpc (43--44 kpc in the lower-resolution cubes), which is about 60 times $R_e$. This is consistent, within its scatter, with the \hi\ size--mass relation of \citet{WangJ2016}, which predicts 35--39 kpc for the measured \hi\ masses. The \hi\ disk has an inclination of $\approx43^\circ$, which we derived from the beam-deconvolved axis ratio of this ellipse assuming a thin disk ($44^\circ$ for an intrinsic axis ratio of 0.2). Its column density has two peaks along the major axis and a depression at the center (Figure~\ref{fig:hi}b). Hence, most of the \hi\ gas resides outside the stellar body of the galaxy, extending well beyond it, with a central low column density region. The Galaxy Evolution Explorer (GALEX) far- and near-ultraviolet images show ultraviolet emission within the optical extent of the galaxy. However, there is no significant ultraviolet emission over the \hi\ disk beyond it.

\subsection{\texorpdfstring{Kinematics of the Stars, the Ionized Gas, and the \hi}{Kinematics of the Stars, the Ionized Gas, and the HI}} \label{sec:kin} \label{sec:stargas}

\begin{figure*}[t!]
\centering
\includegraphics[width=0.85\textwidth]{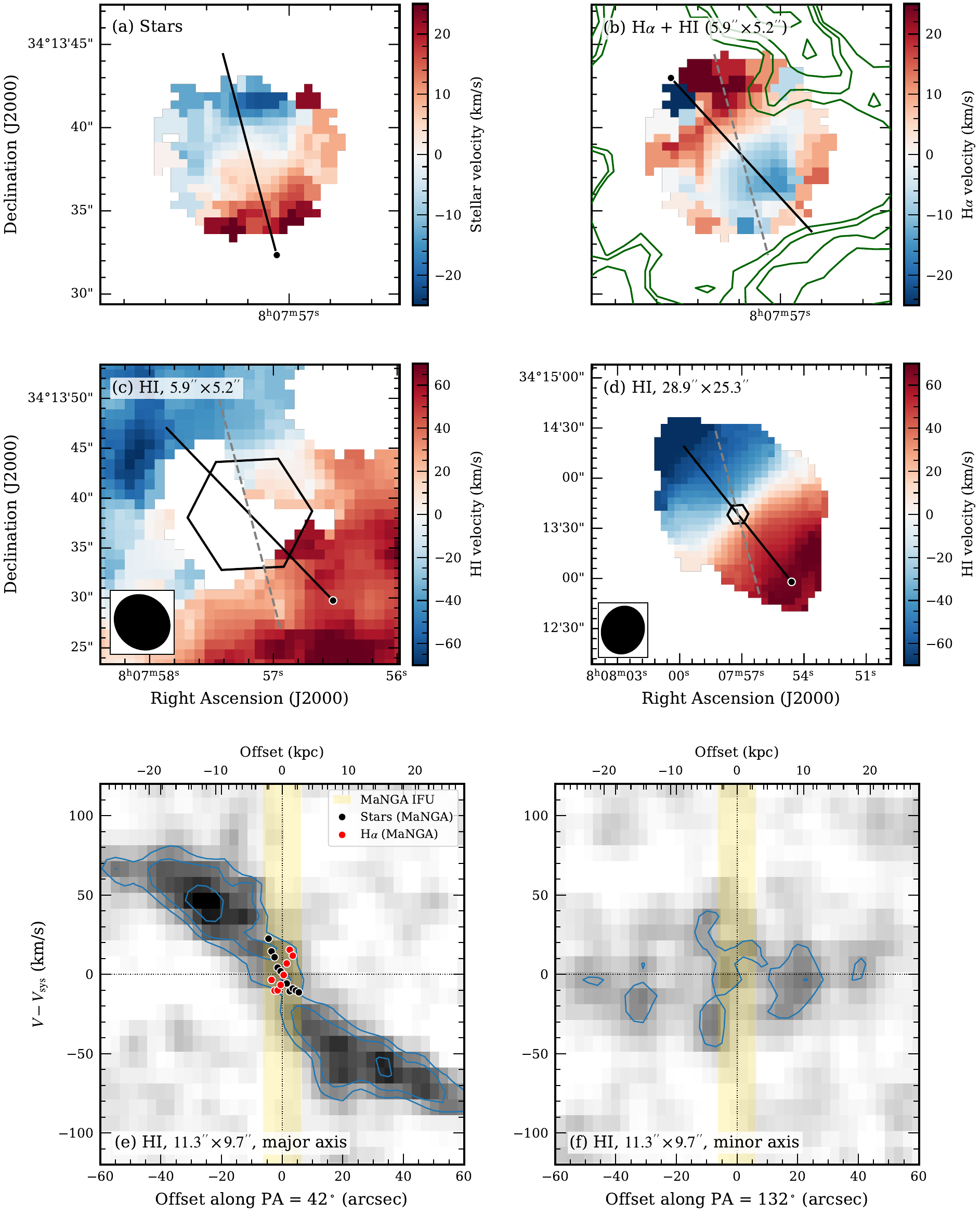}
\caption{Kinematics of the stars, the ionized gas, and the \hi\ of MaNGA~10214-1902. (a) MaNGA stellar and (b) \ha\ velocity maps ($\pm9''$), with the $5.9''\times5.2''$ resolution, green $3\sigma$ ($\rm N_{HI} = 4.6\times 10^{20}\ cm^{-1}$) \hi\ column-density contours increasing by factors of 1.5, which show the \hi\ distribution. (c) \hi\ velocity map from the $5.9''\times5.2''$ cube ($\pm15''$) with the MaNGA footprint (black hexagon), and (d) from the $28.9''\times25.3''$ cube. Solid lines: kinematic major axis of each component (dot on the receding side); dashed gray lines: stellar major axis. Velocities are relative to the systemic velocity of each component. (e), (f) \hi\ PV diagrams of the $11.3''\times9.7''$ cube along the major (PA\,$=42^\circ$; positive offsets toward the receding side of the ionized gas) and minor axes of the ionized gas (slit of 1.5 beams, smoothed to 30\,\kms; contours at $\pm3\sigma\times1.5^n$, where $\sigma$ is the rms noise of each PV diagram). In (e), red and black points are the mean \ha\ (S/N\,$>10$) and stellar velocities in a $4.1''$-wide slit ($1.5\times$ the PSF FWHM), in $1''$ bins; the yellow band marks the MaNGA 19 fiber field of view.}
\label{fig:kin}
\end{figure*}

Figure~\ref{fig:kin} shows the kinematics of the stars, the ionized gas, and the \hi. We measured their global kinematic position angles (PA) using \texttt{fit\_kinematic\_pa} \citep{Krajnovic2006} and listed them in Table~\ref{tab:props}. The stellar velocity map shows only weak rotation ($|V| \lesssim 20$ \kms) compared with the velocity dispersion ($\sigma_* \approx 77$ \kms; Figure~\ref{fig:kin}a), i.e., the stars are dispersion-dominated, as expected for an elliptical galaxy. Within the MaNGA field of view, both the stars and the ionized gas show ordered rotation, but in opposite directions, with $\Delta{\rm PA}({\rm stars}, {\rm H}\alpha) = 152^\circ\pm20^\circ$ ($3\sigma$). The Pipe3D velocity maps also give consistent position angles ($38^\circ$ for \ha\ and $\approx200^\circ$ for the stars). So, the observed counter-rotation does not depend on the analysis pipeline.

The \hi\ disk rotates in the same sense as the stars. On large scales, the \hi\ position angle is $216^\circ$--$221^\circ$ at every resolution (Appendix~\ref{app:hi}), offset from the stellar position angle by $\Delta{\rm PA} = 23^\circ\pm20^\circ$, within the conventional $30^\circ$ threshold for aligned rotation. On the scale of the MaNGA field of view, the $5.9''\times5.2''$ cube resolves the \hi\ that immediately surrounds the ionized gas ($r<12''$, i.e., $\approx5$ kpc; Figure~\ref{fig:kin}c). The kinematic position angle of this \hi\ cube ($224^\circ\pm7^\circ$) differs from that of the ionized gas by $\Delta{\rm PA}({\rm H}\alpha$, \hi) $= 179^\circ\pm10^\circ$. The position--velocity (PV) diagram of the $11.3''\times9.7''$ cube along the kinematic axis of the ionized gas is shown in Figure~\ref{fig:kin}e. This axis lies within $\sim5^\circ$ of the \hi\ major axis, so the same slit samples the full rotation of both components. It shows that, along the same slit, the \hi\ and the stellar velocities decrease toward the receding side of the ionized gas, whereas the \ha\ velocities increase. Along the minor axis, the \hi\ stays close to the systemic velocity (Figure~\ref{fig:kin}f), as expected for a rotating disk. Therefore, the ionized gas clearly counter-rotates with respect to both the stars and the \hi.

However, we must remember that the two gas phases occupy different regions. The ionized gas lies within $r\lesssim2$~kpc, where the \hi\ column density is lowest. In the $5.9''\times5.2''$ cube (beam $\approx2.7$~kpc), \hi\ is detected over only one third of the MaNGA field of view, mostly near its edges (Figure~\ref{fig:kin}b). We therefore compare the \hi\ ($r\approx2$--5~kpc) and the ionized gas ($r\lesssim2$~kpc) at different radii, and we cannot rule out a faint inner \hi\ component, below the sensitivity of our high-resolution cubes (Table~\ref{tab:cubes}), that either co-rotates with the ionized gas or has a different orientation. This produces a three-component kinematic configuration in a single galaxy. Stars and an extended \hi\ disk share the same sense of rotation, and a central ionized gas disk that counter-rotates with respect to both. In the misaligned ETGs of ATLAS$^{3D}$, all gas phases are aligned where they overlap \citep{Davis2011}; here, the \hi\ and the ionized gas occupy different radii, similar to the radial reversals of the gas rotation in NGC~4826 and NGC~4449 \citep{Braun1994, Hunter1999}.

\subsection{Ionization Source and Star Formation Rate} \label{sec:ion}

\begin{figure*}[t!]
\centering
\includegraphics[width=\textwidth]{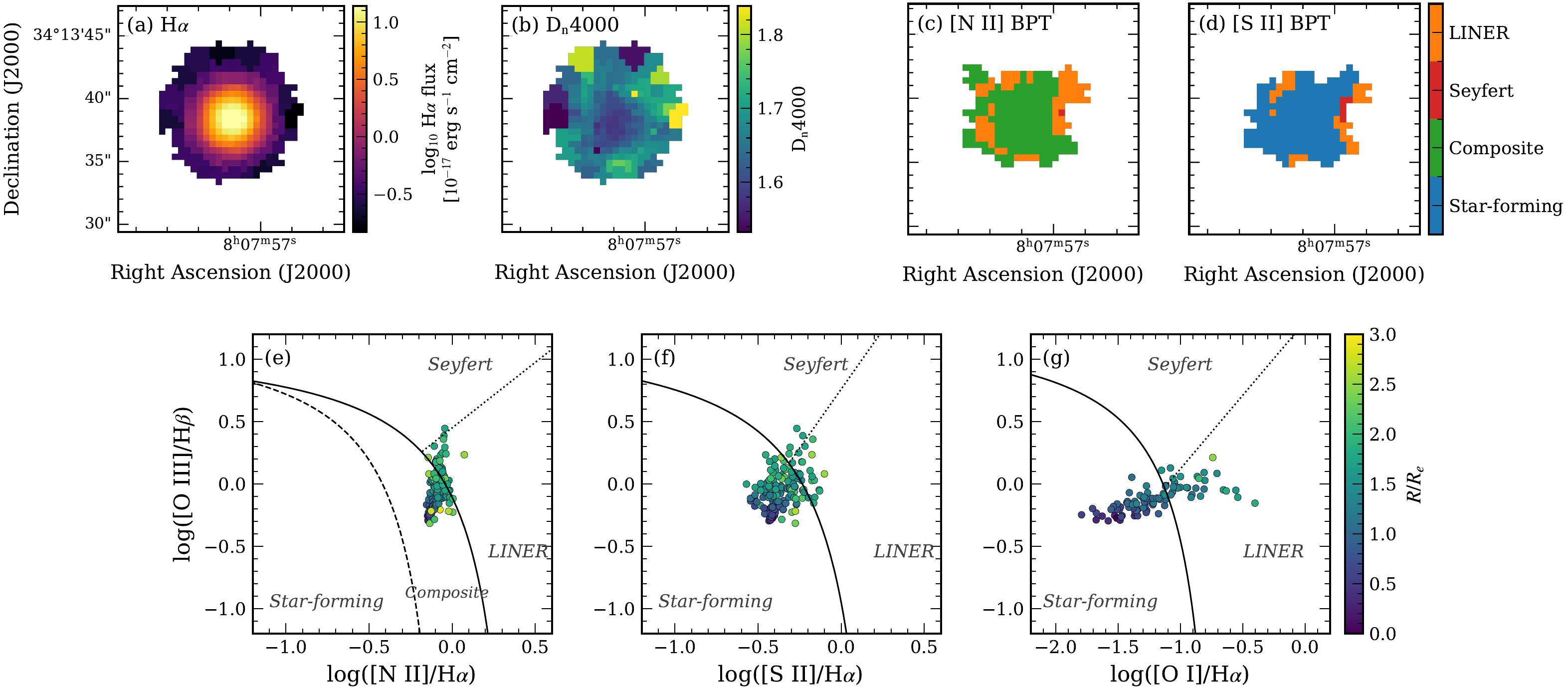}
\caption{Ionization and stellar population of MaNGA~10214-1902 from the MaNGA DAP. Top: (a) \ha\ flux map, (b) D$_{\rm n}$4000 map, and the spaxel classifications in the (c) \nii\ and (d) \sii\ diagnostic diagrams. Bottom: (e) \nii, (f) \sii, and (g) \oi\ diagnostic diagrams for the individual Voronoi bins, colored by $R/R_e$. The solid lines are the maximum-starburst lines of \citet{Kewley2001}, the dashed line is from \citet{Kauffmann2003dec}, and the dotted lines separate the Seyferts from the LINERs \citep{Schawinski2007, Kewley2006}. The ionization regions are labeled in each diagram.}
\label{fig:ion}
\end{figure*}

To identify the source of ionization, we constructed spatially resolved emission-line diagnostic diagrams (Figure~\ref{fig:ion}). In the \nii-based BPT diagram \citep{Baldwin1981, Kewley2001, Kauffmann2003dec}, none of the Voronoi bins are purely star-forming. Of the total bins with S/N\,$>3$ in all the required lines, 75\% are composite, and 24\% are LINER-like. However, in the \sii\ and \oi\ diagrams \citep{Kewley2006}, the central region lies in the star-forming locus: 85\% and 65\% of the bins with S/N\,$>3$ in all the required lines are classified as star-forming, respectively, while the remaining are mostly LINER-like. In all three diagrams, LINER-like spaxels lie at the edge of the field. Hot, low-mass evolved stars can also ionize the gas in ETGs, but they produce EW(\ha)\,$\lesssim3$\,\AA\ \citep{CidFernandes2011}. In the target, the central EW(\ha) is 8.1\,\AA\ and exceeds 6\,\AA\ throughout the PSF-sized region within $R_e$. In contrast, the LINER-like spaxels in the outer region have EW(\ha)\,$\approx3$\,\AA, consistent with diffuse ionized gas \citep{Zhang2017}. Considering all three diagrams together (see also Appendix~\ref{app:optical}), the central ionized gas is likely photoionized mainly by young stars, with a composite \nii/\ha\ ratio, as is often found at the centers of ETGs.

As an independent check for a hidden active galactic nucleus (AGN), we compared the SFR derived from the uGMRT 1.4 GHz continuum (Figure~\ref{fig:hi}d) with the SED-based SFR. Both trace the star formation averaged over the last $\sim100$ Myr \citep{Murphy2011, Salim2016}. The flux density of $284\pm38\,\mu$Jy corresponds to SFR$_{1.4\,{\rm GHz}} = 0.21\pm0.03\,\msun$\,yr$^{-1}$ \citep{Murphy2011}, which agrees with the value of $0.22^{+0.08}_{-0.06}\,\msun$\,yr$^{-1}$ from the Salim catalog. If a significant fraction of the radio emission originated from an AGN, the radio-based SFR would exceed the SED-based value. We find no evidence for a dominant AGN (see also Appendix~\ref{app:gmrt}). The dust-corrected (Balmer decrement; \citealt{Cardelli1989}) \ha\ emission within the MaNGA field of view gives SFR$_{{\rm H}\alpha} = 0.10\,\msun$\,yr$^{-1}$ \citep{Kennicutt2012}, about half of the SED-based value. The gas-phase metallicity is discussed in Appendix~\ref{app:optical}.

\subsection{Stellar Populations} \label{sec:stars}

The target is a star-forming galaxy ($\log{\rm sSFR/yr^{-1}} = -10.55$) following the sSFR-based criterion ($\log{\rm sSFR/yr^{-1}} > -10.8$) of \citet{Salim2014}, although it lies on the lower side of the SFMS. It is a compact ($R_e = 0.78$ kpc) galaxy and has a centrally concentrated light profile with a S\'ersic index of $n_r = 3.6$. Within $R_e$, the Pipe3D light-weighted age is 1.9 Gyr, which is much younger than the mass-weighted age of 3.5 Gyr. As younger stars dominate the light but not the mass, this indicates later star formation on top of an older population. Also, both the ages within $R_e$ are younger than the ages (3.2 and 5.2 Gyr) at $>R_e$ ($3''-6''$), by $\approx0.2$ dex. Given the formal Pipe3D uncertainties ($\approx0.04$ dex within $R_e$ and $\approx0.10$ dex at $3''-6''$), this is significant at only $\approx2\sigma$, but the model-independent D$_{\rm n}$4000 shows the same trend (see below). Stars younger than 0.5 Gyr contribute 7\% of the light within $R_e$, but only 3\% at $3''$--$6''$. However, most of the light comes from intermediate-age (1--5 Gyr) stars. Therefore, the target is not an elliptical galaxy with an old stellar population and a small fraction of young stars, but a galaxy with an extended or episodic star-formation history, with its most recent star formation at its center. These light fractions could, however, be affected by the age--metallicity degeneracy and should be treated qualitatively. Also, as $R_e$ is smaller than the PSF, the central values are averages over $r\lesssim1.5''$, and the radial differences are lower limits to the intrinsic ones.

The DAP maps, which do not depend on the stellar population modeling, show the same picture. The D$_{\rm n}$4000 index, which increases with the stellar age, is lowest at the center ($1.60\pm0.01$) and increases outward ($1.67\pm0.07$ at $3''$--$6''$; median and scatter of the Voronoi bins; Figure~\ref{fig:ion}b). A central value of 1.60 is close to the D$_{\rm n}$4000\,$\approx1.5$ that separates the galaxies dominated by young and old stellar populations \citep{Kauffmann2003may}, so the central light is dominated by intermediate-age stars or a mixture of old and young stars (See also Appendix \ref{app:optical}). Further, the \ha\ equivalent width, which measures the current star formation relative to the existing stellar population, exceeds 6 \AA\ at the center, well above the $\lesssim3$ \AA\ that old stars alone can produce (Section~\ref{sec:ion}). These results indicate that the center hosts a younger stellar population than the outskirts and is still forming stars.

\section{Discussion} \label{sec:discussion}

Counter-rotation in galaxies is observed in three forms: gas counter-rotating with respect to the stars \citep{Bertola1992, Davis2011, Chen2016}, two counter-rotating stellar disks \citep{Rubin1992, Corsini2014, Bevacqua2022, Bao2026}, and nested, counter-rotating or misaligned gaseous disks, in which an inner gas disk rotates in a different direction from an outer gas disk that surrounds it \citep{Braun1994, Hunter1999, Serra2014, cao2022Nat}. MaNGA~10214-1902 belongs to the last, and rare, class. Unlike the dual misaligned gas disks of \citet{cao2022Nat} found in MaNGA, where both ionized gas disks are misaligned with the stars and require two acquisition events, here the \hi\ co-rotates with the stars, so only the counter-rotating ionized gas requires an external origin. Gas lost by the stars of the galaxy itself (e.g., in the winds of asymptotic giant branch stars, planetary nebulae, and supernovae) keeps the angular momentum of its parent stars, so this internal gas would co-rotate with the stars \citep{Sarzi2006, Davis2011}. So, the counter-rotating ionized gas must have been acquired from outside, either through a gas-rich minor or dwarf merger or by accretion from the surrounding medium with opposite angular momentum, as seen in simulations \citep{Starkenburg2019, Khim2021, Casanueva2022}. The tidal features identified in the visual morphology \citep{VazquezMata2022} may suggest a past interaction. Within the GBT beam, the blueberry galaxy J0807+3414 (Appendix~\ref{app:j0807}) lies at a similar redshift. It is a low-mass ($\log M_*/\msun = 6.2$) starburst $62''$ (28 kpc) north of the target with a velocity offset of $\approx180$\,\kms, which is probably a satellite. It is not detected in \hi\ ($\log M_{\rm HI}/\msun < 8.1$).

A strongly warped gas disk can also mimic counter-rotation in projection. However, a warp cannot reverse the direction of rotation along the major axis \citep{cao2022Nat}. In the target, the \hi\ around the ionized gas and the \ha\ show velocity gradients of opposite sign along the same slit in the major-axis PV diagram (Section~\ref{sec:kin}). Further, the \hi\ position angle changes by $\lesssim16^\circ$ between the inner $12''$ and the edge of the disk at all resolutions, and the $\approx180^\circ$ change between the ionized gas and the \hi\ coincides with the central \hi\ depression. The \hi\ disk of the target shows no strong warp, and the reversal occurs between the \hi\ and the ionized gas. However, as the \hi\ co-spatial with the ionized gas is not detected, a strong warp in an inner \hi\ component cannot be fully ruled out.

The \hi\ disk co-rotates with the stars, which suggests that it is either a long-lived reservoir or was acquired earlier with the same angular momentum as the stars. On the other hand, the ionized gas has the opposite angular momentum and must have been acquired later. Two counter-rotating gas components cannot coexist at the same radius for long; where they meet, the collisions between gas clouds are expected to cancel their angular momentum and drive the gas toward the center \citep{Chen2016, Zhou2022}. With an \hi\ rotation velocity of $\approx103$\,\kms\ ($W_{50}/2\sin i$), the orbital period is $\approx60$\,Myr at 1\,kpc and $\approx150$\,Myr at 2.5\,kpc. The nested configuration should therefore survive only for a few orbital periods ($\lesssim$ a few hundred Myr), consistent with the timescale estimated for dual misaligned gas disks \citep{cao2022Nat} and with its rarity. This lifetime is much shorter than the $\sim1$--2 Gyr over which the central intermediate-age population formed (Section~\ref{sec:stars}). Therefore, the counter-rotating gas can fuel only the most recent star formation, i.e., the current SFR and the stars younger than $\sim0.5$ Gyr (7\% of the light within $R_e$), while the intermediate-age population must have formed from earlier gas, either co-rotating gas from the \hi\ disk or previous accretion events. This suggests that the gas acquisition in the target could be episodic. The central \hi\ depression may be related to such an inflow, but central \hi\ depressions are also common in disk galaxies where the gas is mostly molecular, which CO observations can test.

If the accreted gas loses its angular momentum, it can fuel the most recent central star formation of the target, i.e., the high EW(\ha) and the ongoing star formation (Sections~\ref{sec:ion} and \ref{sec:stars}). \citet{cao2022Nat} found enhanced star formation at the interface of two misaligned gas disks. In MaNGA~10214-1902, the star formation peaks at the center, inside the counter-rotating ionized gas disk. The inflowing gas can also feed the central black hole \citep{Tous2026}, although we do not find evidence for an AGN in the target. A similar central growth is seen in the star-forming counter-rotators of MaNGA \citep{Chen2016}, and \citet{Tous2026} showed that misaligned accretion can sustain main-sequence SFRs only below $M_* \sim 10^{10}\,\msun$, where the accreted gas is not diluted by a massive pre-existing bulge. MaNGA~10214-1902, with $M_* = 10^{9.9}\,\msun$, lies at the edge of this regime. Together with the kinematically disturbed velocity maps of most SF-Es \citep{Biswas2024}, this suggests that the accretion of misaligned gas may be one of the channels through which low-mass ellipticals sustain their star formation. At the current SFR, the counter-rotating gas forms $\lesssim6\times10^7\,\msun$ of stars in a few hundred Myr, $<1\%$ of the stellar mass, which is too small to affect the stellar kinematics measured here. If the accretion continues, however, it may build a young, counter-rotating stellar disk \citep{Bevacqua2022, Bao2026}, and once the central gas is consumed, the star formation should fade, leaving the co-rotating \hi\ disk as a low-column-density reservoir, as seen in many \hi-rich ETGs \citep{Serra2012, Serra2014}. With an \hi\ depletion time of $M_{\rm HI}/{\rm SFR}\approx16$--20 Gyr, much longer than the $\sim3$ Gyr typical of star-forming galaxies \citep{Schiminovich2010}, most of the \hi\ disk is not forming stars efficiently, consistent with the absence of ultraviolet emission over the disk (Section~\ref{sec:hi}).

\section{Summary and Conclusions} \label{sec:summary}

In this Letter, we have combined deep uGMRT \hi\ imaging with SDSS-MaNGA spatially resolved spectroscopy to understand where the gas that fuels the star formation in the elliptical galaxy MaNGA~10214-1902 comes from. We find that the galaxy hosts a large reservoir of atomic gas, $\log(M_{\rm HI}/\msun) = 9.65$, in a 45 kpc \hi\ disk that rotates in the same sense as the stars. However, at the center of this disk, where the \hi\ column density is the lowest, the ionized gas rotates in the opposite direction to both the stars and the surrounding \hi. To our knowledge, this is the first ETG in which the ionized gas at the center of a flat \hi\ disk that co-rotates with the stars counter-rotates with respect to both (the two phases are compared at different radii; Section~\ref{sec:kin}). The central gas is photoionized primarily by young stars, and the agreement between the radio- and SED-based SFRs provides no evidence of a dominant AGN. The center hosts a younger stellar population than the outskirts and is still forming stars. These results are consistent with the acquisition of gas with opposite angular momentum into the center of a pre-existing, co-rotating \hi\ disk, where collisions between the two gas components are expected to remove angular momentum and drive the gas inward to fuel the most recent central star formation. Such a configuration should last only a few hundred Myr, which may explain its rarity. The acquisition of misaligned gas may therefore be one of the channels through which low-mass elliptical galaxies sustain their star formation. CO observations, deeper \hi\ imaging of the central region, and resolved \hi\ imaging of a sample of SF-Es will test this picture and establish how common such configurations are.

\begin{acknowledgments}
PB, YW, and SK acknowledge the support of the Department of Atomic Energy, Government of India, under the project title ``Next Generation Instrumentation for Radio Astronomy'', UID number RTI4017. OB was supported by the National Science Foundation under Cooperative Agreement 2421782 and the Simons Foundation grant MPS-AI-00010515 awarded to the NSF-Simons AI Institute for Cosmic Origins — CosmicAI, https://www.cosmicai.org/.

We thank the staff of the GMRT that made these observations possible. The GMRT is run by the National Centre for Radio Astrophysics of the Tata Institute of Fundamental Research. The Green Bank Observatory is a facility of the National Science Foundation operated under cooperative agreement by Associated Universities, Inc.

Funding for the Sloan Digital Sky Survey IV has been provided by the Alfred P. Sloan Foundation, the U.S. Department of Energy Office of Science, and the Participating Institutions. SDSS-IV acknowledges support and resources from the Center for High-Performance Computing at the University of Utah. The SDSS website is www.sdss.org. SDSS-IV is managed by the Astrophysical Research Consortium for the Participating Institutions of the SDSS Collaboration, including the Brazilian Participation Group, the Carnegie Institution for Science, Carnegie Mellon University, Center for Astrophysics-Harvard \& Smithsonian, the Chilean Participation Group, the French Participation Group, Instituto de Astrof\'isica de Canarias, The Johns Hopkins University, Kavli Institute for the Physics and Mathematics of the Universe (IPMU) / University of Tokyo, the Korean Participation Group, Lawrence Berkeley National Laboratory, Leibniz Institut f\"ur Astrophysik Potsdam (AIP),  Max-Planck-Institut f\"ur Astronomie (MPIA Heidelberg), Max-Planck-Institut f\"ur Astrophysik (MPA Garching), Max-Planck-Institut f\"ur Extraterrestrische Physik (MPE), National Astronomical Observatories of China, New Mexico State University, New York University, University of Notre Dame, Observat\'ario Nacional / MCTI, The Ohio State University, Pennsylvania State University, Shanghai Astronomical Observatory, United Kingdom Participation Group, Universidad Nacional Aut\'onoma de M\'exico, University of Arizona, University of Colorado Boulder, the University of Oxford, University of Portsmouth, University of Utah, University of Virginia, University of Washington, University of Wisconsin, Vanderbilt University, and Yale University. 

This work uses the MaNGA DAP, Pipe3D, PyMorph, deep learning and visual morphology, HI-MaNGA value-added catalogs, and the GALEX-SDSS-WISE Legacy Catalog. 

The Legacy Surveys consist of three individual and complementary projects: the Dark Energy Camera Legacy Survey (DECaLS; Proposal ID \#2014B-0404; PIs: David Schlegel and Arjun Dey), the Beijing-Arizona Sky Survey (BASS; NOAO Prop. ID \#2015A-0801; PIs: Zhou Xu and Xiaohui Fan), and the Mayall z-band Legacy Survey (MzLS; Prop. ID \#2016A-0453; PI: Arjun Dey). DECaLS, BASS and MzLS together include data obtained, respectively, at the Blanco telescope, Cerro Tololo Inter-American Observatory, NSF’s NOIRLab; the Bok telescope, Steward Observatory, University of Arizona; and the Mayall telescope, Kitt Peak National Observatory, NOIRLab. The Legacy Surveys project is honored to be permitted to conduct astronomical research on Iolkam Du’ag (Kitt Peak), a mountain with particular significance to the Tohono O’odham Nation.

NOIRLab is operated by the Association of Universities for Research in Astronomy (AURA) under a cooperative agreement with the National Science Foundation.

This project used data obtained with the Dark Energy Camera (DECam), which was constructed by the Dark Energy Survey (DES) collaboration. Funding for the DES Projects has been provided by the U.S. Department of Energy, the U.S. National Science Foundation, the Ministry of Science and Education of Spain, the Science and Technology Facilities Council of the United Kingdom, the Higher Education Funding Council for England, the National Center for Supercomputing Applications at the University of Illinois at Urbana-Champaign, the Kavli Institute of Cosmological Physics at the University of Chicago, Center for Cosmology and Astro-Particle Physics at the Ohio State University, the Mitchell Institute for Fundamental Physics and Astronomy at Texas A\&M University, Financiadora de Estudos e Projetos, Fundacao Carlos Chagas Filho de Amparo a Pesquisa do Estado do Rio de Janeiro, Conselho Nacional de Desenvolvimento Cientifico e Tecnologico and the Ministerio da Ciencia, Tecnologia e Inovacao, the Deutsche Forschungsgemeinschaft and the Collaborating Institutions in the Dark Energy Survey. The Collaborating Institutions are Argonne National Laboratory, the University of California at Santa Cruz, the University of Cambridge, Centro de Investigaciones Energeticas, Medioambientales y Tecnologicas-Madrid, the University of Chicago, University College London, the DES-Brazil Consortium, the University of Edinburgh, the Eidgenossische Technische Hochschule (ETH) Zurich, Fermi National Accelerator Laboratory, the University of Illinois at Urbana-Champaign, the Institut de Ciencies de l’Espai (IEEC/CSIC), the Institut de Fisica d’Altes Energies, Lawrence Berkeley National Laboratory, the Ludwig Maximilians Universitat Munchen and the associated Excellence Cluster Universe, the University of Michigan, NSF’s NOIRLab, the University of Nottingham, the Ohio State University, the University of Pennsylvania, the University of Portsmouth, SLAC National Accelerator Laboratory, Stanford University, the University of Sussex, and Texas A\&M University.

BASS is a key project of the Telescope Access Program (TAP), which has been funded by the National Astronomical Observatories of China, the Chinese Academy of Sciences (the Strategic Priority Research Program “The Emergence of Cosmological Structures” Grant \# XDB09000000), and the Special Fund for Astronomy from the Ministry of Finance. The BASS is also supported by the External Cooperation Program of the Chinese Academy of Sciences (Grant \# 114A11KYSB20160057), and the Chinese National Natural Science Foundation (Grant \# 11433005).
The Legacy Survey team makes use of data products from the Near-Earth Object Wide-field Infrared Survey Explorer (NEOWISE), which is a project of the Jet Propulsion Laboratory/California Institute of Technology. NEOWISE is funded by the National Aeronautics and Space Administration. The Legacy Surveys imaging of the DESI footprint is supported by the Director, Office of Science, Office of High Energy Physics of the U.S. Department of Energy under Contract No. DE-AC02-05CH1123, by the National Energy Research Scientific Computing Center, a DOE Office of Science User Facility under the same contract, and by the U.S. National Science Foundation, Division of Astronomical Sciences under Contract No. AST-0950945 to NOAO.
\end{acknowledgments}


\facilities{GMRT, GBT, Sloan, Blanco (DECam)}

\software{CASA \citep{casa},
          SoFiA-2 \citep{Serra2015, Westmeier2021},
          astropy \citep{Astropy2013, Astropy2018, Astropy2022},
          numpy \citep{numpy},
          scipy \citep{scipy},
          \texttt{fit\_kinematic\_pa} \citep{Krajnovic2006},
          matplotlib \citep{matplotlib},
          aplpy \citep{aplpy2012, aplpy2019}}

\appendix

\section{uGMRT Data Reduction and Radio Continuum} \label{app:gmrt}

The resolved \hi\ 21 cm observation was carried out with the uGMRT Band~5 receiver with a 100 MHz bandwidth, centered at 1400 MHz and split into 8192 channels (proposal 47\_127, PI: P.~Biswas). This gives a native velocity resolution of $\approx2.6$\,\kms\ at the redshifted \hi\ frequency. 3C48 and 3C286 were used to determine the flux density scale and the bandpass, and the compact source 0741+312, observed for 5 minutes after every 45 minutes on the target, was used to solve for the antenna-based complex gains.

We used \texttt{CASA} to analyze the data and followed the standard procedure. First, the nonworking antennas and the data affected by radio-frequency interference (RFI) were flagged. The antenna-based complex gains and bandpass were then measured using the tasks \texttt{GAINCAL} and \texttt{BANDPASS} and applied to the target data, which were further flagged using a combination of \texttt{TFCROP} and \texttt{RFLAG}. A line-free continuum data set, averaged over 15 channels, was then self-calibrated using a few rounds of phase-only and amplitude-and-phase self-calibration and imaging. The imaging was done using multi-term multi-frequency synthesis \citep{Rau2011}, $w$-projection \citep{Cornwell2008}, and Briggs \citep{Briggs1995} weighting with a robust parameter of 0. The self-calibration gains were then applied to the full-resolution data.

For each of the six resolutions, a continuum model was made at matching resolution (using the same $uv$-range and tapering) and subtracted from the self-calibrated data in the $uv$-plane using the task \texttt{UVSUB}. The continuum-subtracted visibilities were then imaged into cubes at a velocity resolution of 10\,\kms\ using Briggs weighting with a robust parameter of 0.5, selecting the $uv$-ranges listed in Table~\ref{tab:cubes} (with additional $uv$-tapering), and cleaned with the H\"ogbom algorithm down to $\sim0.5\sigma$. A final continuum was then estimated in the line-free channels and subtracted from each cube using the task \texttt{IMCONTSUB} with a \texttt{FITORDER} of 2. The \hi\ sources were identified with SoFiA-2 using the smooth and clip algorithm, with spatial Gaussian kernels of 0, 4, and 8 pixels and spectral boxcar kernels of 0, 3, 7, and 15 channels, a threshold of 3.5--4$\sigma$, and a reliability cut of 90\% \citep{SerraJurek2012}. We calculated the \hi\ masses as $M_{\rm HI} = 2.356\times10^5\,(1+z)^{-2}\,D_L^2\,S$, where $D_L$ is in Mpc and the integrated flux $S$ in Jy\,\kms, following the HI-MaNGA convention \citep{Meyer2017, Stark2021}.

The self-calibrated continuum image (Figure~\ref{fig:hi}d) has a resolution of $2.3''\times1.9''$ and an rms noise of $9\,\mu$Jy\,beam$^{-1}$. The galaxy is detected at the optical position with a peak flux density of $195\,\mu$Jy\,beam$^{-1}$ ($\approx22\sigma$). A two-dimensional Gaussian fit with the \texttt{CASA} task \texttt{IMFIT} gives an integrated flux density of $284\pm25\,\mu$Jy. Including a 10\% uncertainty in the GMRT flux density scale, added in quadrature, the total uncertainty is $\pm38\,\mu$Jy. This corresponds to $L_{1.4\,{\rm GHz}} = 3.2\times10^{20}$\,W\,Hz$^{-1}$ ($K$-corrected assuming a spectral index of $-0.7$) and, using the calibration of \citet{Murphy2011}, to SFR$_{1.4\,{\rm GHz}} = 0.21\pm0.03\,\msun$\,yr$^{-1}$. Here, we note that the radio-based SFR calibrations may underestimate the SFR of low-luminosity galaxies \citep{Bell2003}. Hence, the agreement between the 1.4 GHz radio continuum-based SFR and the SED-based SFR value from the Salim catalog indicates that there is no radio-bright AGN present in this galaxy.

\section{\texorpdfstring{\hi\ at Multiple Resolutions}{HI at Multiple Resolutions}} \label{app:hi}

\begin{figure*}[h!]
\centering
\includegraphics[width=0.95\textwidth]{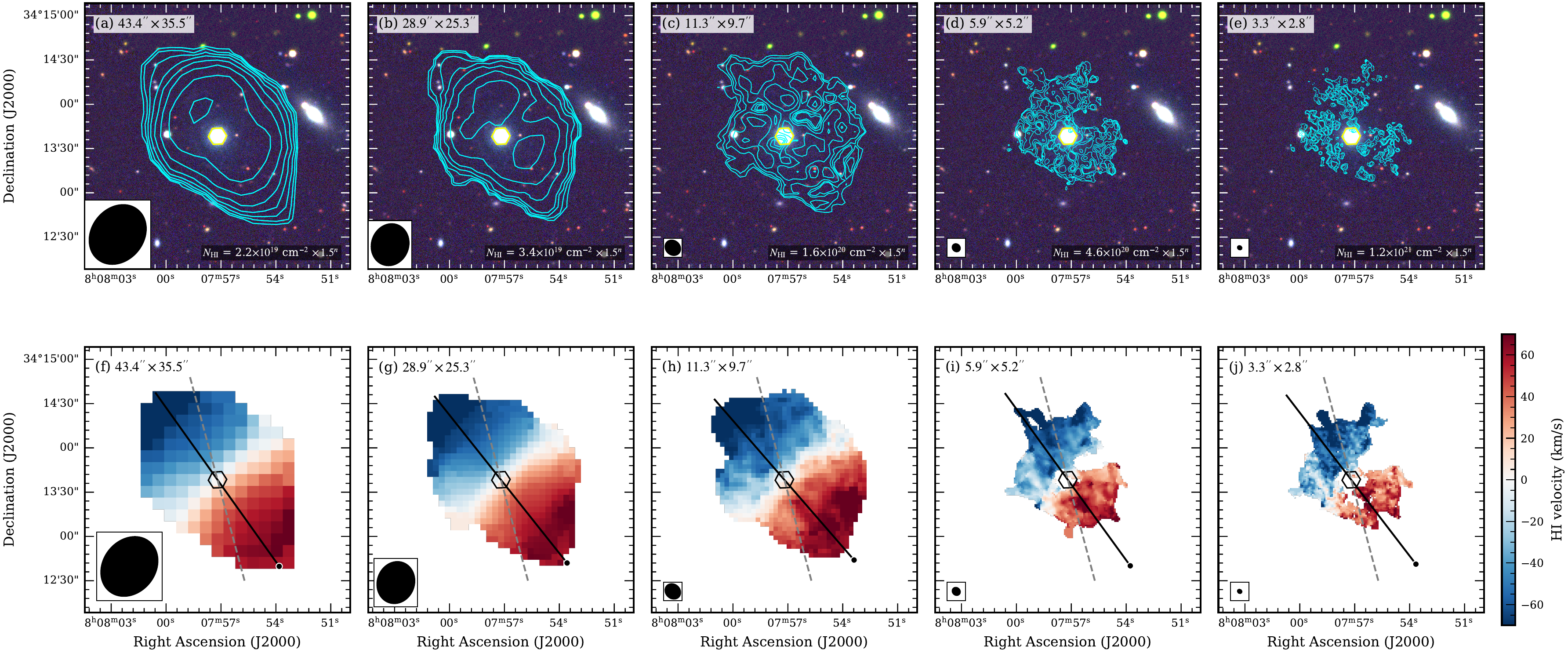}
\caption{uGMRT \hi\ at five of the six angular resolutions (Table~\ref{tab:cubes}; the $40.2''\times31.1''$ cube, which is very similar to the $43.4''\times35.5''$ cube, is not shown) on a common $\pm90''$ field. (a)--(e) DESI Legacy Surveys $grz$ images with the \hi\ column-density contours (cyan). The contours start at $3\sigma$ over one channel and increase by factors of 1.5; the lowest level is given in each panel. (f)--(j) the corresponding \hi\ velocity maps. The solid lines show the \hi\ kinematic major axis (with a dot on the receding side), and the dashed lines show the stellar major axis. The hexagons (yellow in (a)--(e) and black in (f)--(j)) mark the MaNGA IFU footprint, and the ellipses in the lower-left corners show the beams.}
\label{fig:hires}
\end{figure*}

Table~\ref{tab:cubes} lists the parameters of the \hi\ spectral cubes, and Figure~\ref{fig:hires} shows the \hi\ column-density distribution and velocity maps at five of them. The kinematic position angle of the \hi\ disk is $216^\circ$--$221^\circ$ in every cube. So, the co-rotation of the \hi\ with the stars does not depend on the resolution.

\begin{deluxetable}{ccccc}
\tablecaption{Parameters of the \hi\ Spectral Cubes \label{tab:cubes}}
\tablehead{\colhead{$uv$-range} & \colhead{Beam} & \colhead{Beam} & \colhead{rms} & \colhead{$N_{\rm HI}$ ($3\sigma$)} \\
\colhead{(k$\lambda$)} & \colhead{($''\times''$)} & \colhead{(kpc $\times$ kpc)} & \colhead{(mJy\,beam$^{-1}$)} & \colhead{(cm$^{-2}$)}}
\startdata
0.5--80 & $3.26\times2.80$   & $1.49\times1.28$   & 0.34 & $1.24\times10^{21}$ \\
0.5--40 & $5.88\times5.21$   & $2.68\times2.37$   & 0.42 & $4.60\times10^{20}$ \\
0.5--20 & $11.27\times9.73$  & $5.13\times4.43$   & 0.51 & $1.55\times10^{20}$ \\
0.5--10 & $28.87\times25.30$ & $13.15\times11.52$ & 0.76 & $3.44\times10^{19}$ \\
0.5--6  & $40.17\times31.13$ & $18.30\times14.18$ & 1.00 & $2.65\times10^{19}$ \\
0.5--5  & $43.37\times35.50$ & $19.75\times16.17$ & 1.00 & $2.16\times10^{19}$ \\
\enddata
\tablecomments{The columns are (1) $uv$-range used for the cube, (2) angular resolution, (3) corresponding spatial resolution at $z = 0.02178$, (4) rms noise per 10\,\kms\ channel, and (5) the corresponding $3\sigma$ \hi\ column density sensitivity over one channel. The rms noise is lowest in the highest-resolution cubes, as they include more baselines.}
\end{deluxetable}

\section{\texorpdfstring{Additional Optical Measurements}{Additional Optical Measurements}} \label{app:optical}

The integrated line ratios at the center of the target are $\log(\nii/{\rm H}\alpha) = -0.13$, $\log(\sii/{\rm H}\alpha) = -0.42$, $\log(\oi/{\rm H}\alpha) = -1.46$, and $\log(\oiii/{\rm H}\beta) = -0.23$. With EW(\ha)\,$>6$\,\AA\ and $\log(\nii/{\rm H}\alpha) = -0.13$, the center would be classified as a strong AGN in the WHAN diagram of \citet{CidFernandes2011}, which uses only the \nii/\ha\ ratio. However, the \sii\ and \oi\ diagrams and the agreement between the radio and SED-based SFRs (Section~\ref{sec:ion}) favor stars as the main source of ionization. The O3N2 gas-phase metallicity \citep{Pettini2004} of the star-forming spaxels, $12+\log({\rm O/H}) \approx 8.75$, does not show any decrease at the center. Hence, the emission lines do not provide evidence that the acquired gas is metal-poor. This may be due to the mixing of the accreted gas with the pre-existing metal-rich gas and rapid enrichment by new stars \citep[similar to the galaxies in][]{cao2022Nat}. Finally, the SDSS $3''$ fiber spectrum of the nucleus shows strong stellar absorption features (Ca H\&K, Ca G band, Mg b, and Na D) together with narrow Balmer and forbidden emission lines, consistent with the mixed stellar population and the ongoing star formation inferred from the MaNGA data. The SDSS pipeline fits to the emission lines give Gaussian widths of $\sigma\approx92$\,\kms\ for the Balmer lines and $\approx113$\,\kms\ for the forbidden lines, i.e., all the lines are narrow, and there is no evidence for a broad component limiting the presence of a strong AGN.

\section{A Serendipitous Radio Detection of the Blueberry Galaxy J0807+3414} \label{app:j0807}

The field of view of our uGMRT observations contains J0807+3414 (RA = 08$^{\rm h}$07$^{\rm m}$58.0$^{\rm s}$, Dec = +34$^\circ$14$'$39.2$''$), a compact, low-metallicity ($12+\log$[O/H] $< 8.0$) starburst of the ``blueberry'' class at $z = 0.0224$ ($D_L \approx 101.1$~Mpc), with $\log(M_\star/M_\odot) = 6.2$ and SFR$_{\mathrm{H}\alpha} = 0.21~M_\odot$~yr$^{-1}$ \citep[][]{Kouroumpatzakis24}. We detect it at 1.4~GHz in radio continuum with $S_{1.4} = 101.7 \pm 9.0~\mu$Jy (peak $\approx 7\sigma$). Including a 10\% uncertainty in the GMRT flux density scale, added in quadrature, the total uncertainty is $\pm13.6~\mu$Jy. This flux density corresponds to $L_{1.4\,\rm GHz} \approx 1.24 \times 10^{20}$~W~Hz$^{-1}$ ($K$-corrected assuming $\alpha = -0.7$).

The radio-based SFR (SFR$_{\rm radio})$ is $0.08\pm0.01~M_\odot$~yr$^{-1}$ \citep{Murphy2011}, a factor of $\sim$3 below SFR$_{\mathrm{H}\alpha}$, consistent with previous radio observations of blueberry galaxies \citep{Sebastian19,Borkar24-BB-LOFAR, Bait26}. Such radio deficits are common in low-mass galaxies and can be due to escape of cosmic-ray electrons \citep{Bell2003}, young stellar ages \citep{Sebastian19} and/or the effect of free--free absorption due to a high emission measure \citep{Bait26}.

J0807+3414 lies $61.6''$ (28.1\,kpc) north of MaNGA~10214-1902, with a velocity offset of $\approx180$\,\kms\ ($cz = 6729 \rm \ km \ s^{-1}$), and is probably a satellite of the target. We extracted the \hi\ spectrum from the $28.9''\times25.3''$ cube at the position of its radio continuum peak. The spectrum has an rms noise of 0.74\,mJy per 10\,\kms\ channel. It is not detected in \hi\ ($\log M_{\rm HI}/\msun < 8.1$; $3\sigma$ in the $28.9''\times25.3''$ cube for a 50\,\kms\ line width). With $\log M_\star/\msun = 6.2$, the uGMRT limit corresponds to $M_{\rm HI}/M_\star \lesssim 70$ and is consistent with the upper limit on the $M_{\rm HI}/M_\star$ ratio of green pea galaxies at such low stellar masses \citep{Khasnovis25}.

\clearpage
\bibliography{sample701}{}
\bibliographystyle{aasjournalv7}

\end{document}